\documentclass[twocolumn]{aastex63}
\usepackage{graphicx}
\usepackage{amsmath}   
\usepackage{amssymb}

\usepackage{graphicx}  
\usepackage{amsmath}   
\usepackage{amssymb}  
\usepackage{color}

\usepackage{natbib}

\usepackage[active]{srcltx}

\def\eq#1{{Eq.~(\ref{#1})}}

\begin{document}

\title{A new limit on intergalactic magnetic fields on sub-kpc scales from fast radio bursts}

\author{Hamsa Padmanabhan}
\affiliation{ D\'epartement de Physique Th\'eorique, Universit\'e de Gen\`eve \\
24 quai Ernest-Ansermet, CH 1211 Gen\`eve 4, Switzerland\\
}

\author{Abraham Loeb}
\affiliation{Astronomy department, Harvard University \\
60 Garden Street, Cambridge, MA 02138, USA}

\email{hamsa.padmanabhan@unige.ch, aloeb@cfa.harvard.edu}

  \begin{abstract}
 We use the measured scattering timescales of Fast Radio Bursts  (FRBs) from the CHIME catalog to derive an upper limit on the magnetic field on sub-kpc scales in the intergalactic medium (IGM). A nonmagnetized, photoionized IGM is insufficient to explain the turbulent scattering at all redshifts, with a Warm-Hot component being marginally consistent with the data at $z \sim 1$.  Accounting for the lower envelope of the temporal smearing distribution with a nonzero magnetic field leads to upper limits {$B < 10-30$ nG on scales of 0.07-0.20 kpc in the  IGM at  $z \sim 1-2$}. Our work introduces a novel technique to constrain small-scale magnetic fields in the IGM, in a regime unexplored by the Rotation and Dispersion Measures of FRBs.

 \end{abstract}


\section{Introduction}
Magnetic fields are ubiquitous on all scales in the Universe. 
Various constraints have been placed on the strength of magnetic fields in the intergalactic medium  (IGM) from theory and observations \citep[for reviews, see, e.g.,][]{vallee2004,  Kahniashvili2005, durrer2013, subramanian2016}. To date, the strongest constraints are on cosmological scales, with values of up to a few nanoGauss (nG) expected from primordial magnetic fields in the early Universe \citep[][]{Quashnock1988, pshirkov2016, planck2016, cheng1996, kawasaki2012} out to horizon scales and redshifts $z \gtrsim 5$. 

On scales of order a few Mpc, the magnetic field has been measured in filaments of the Warm-Hot Intergalactic Medium \citep[WHIM;][]{dave2001, cen2006}, which plays a key role in structure formation \citep[e.g.,][]{vazza2014}. Constraints on the magnetic field in this phase of the IGM have traditionally been placed by using the Faraday Rotation Measure (RM) from background polarized electromagnetic sources \citep[e.g.,][]{vernstrom2017, vernstrom2019, vernstrom2021, osullivan2019, osullivan2020, locatelli2021}, finding upper limits  of 30-300  nG, consistently with the results of simulations \citep{dolag1999, bruggen2005, ryu2008, vazza2017}. Quasar outflows may pollute the IGM with magnetic fields of the order of 1 nG on scales of $\sim 1$ Mpc \citep{furlanetto2001} by $z \sim 3$. On smaller scales, such as within the Milky Way and in galaxy clusters, the magnetic fields are of order a few $\mu$G \citep[e.g.,][]{beck1996, bernet2008}.

Turbulence in the IGM plays an important role in the amplification of seed magnetic fields \citep[][]{ryu2008, xu2020, macquart2013}. 
 Transient electromagnetic events such as fast radio bursts \citep[FRBs;][]{lorimer2007} can be used to measure the turbulence  in the IGM via  their individual temporal broadening and statistical fluctuations in their Dispersion Measures \citep[DMs;][]{thornton2013, macquart2013, petroff2016, xu2020}. FRBs have been also used to constrain several properties of the WHIM \citep{macquart2020} including its magnetic field, with upper limits measured from the RM distribution being 20-40 nG on scales of 0.5 Mpc to 1 Gpc at $z \lesssim 1$ \citep[][]{ravi2016}. Simulations suggest that a sample of $\gtrsim 10^3$ FRBs with RM $> 1$ rad m$^{-2}$ is required to improve these constraints by an order of magnitude \citep{hackstein2020}. 

 In this paper, we introduce a new technique using the measured scattering timescales of FRBs from the CHIME catalog to place upper limits on  magnetic fields on sub-kpc scales in the IGM. We use the lower envelope of the observed scattering timescale distribution at a given frequency to constrain  the lower scale of turbulence (denoted by $l_0$) in terms of the IGM properties. We find that an unmagnetized, photoionized IGM at $z \lesssim 2$ with $l_0$ given by the Coulomb mean-free path of the plasma is insufficient to explain the observed smearing over $0 < z < 2$. If a Warm-Hot component of the IGM is included, the expected smearing timescales are marginally consistent with the observations at $z \sim 1$, but fall short by 1-2 orders of magnitude at lower redshifts. Accounting for the smearing by a nonzero magnetic field leads to upper limits of $\lesssim 10$ nG at $z \lesssim 1$ at the relevant lower scale for the turbulence (which is given by the viscous scale of the field) of $l_0 \sim 0.01 - 0.2$ kpc.
 Our analysis introduces a new technique to measure magnetic fields on small scales unexplored by approaches that use the Faraday Rotation and Dispersion Measures.

\section{Methods and Results}
\begin{figure}
    \centering
    \includegraphics[width = \columnwidth]{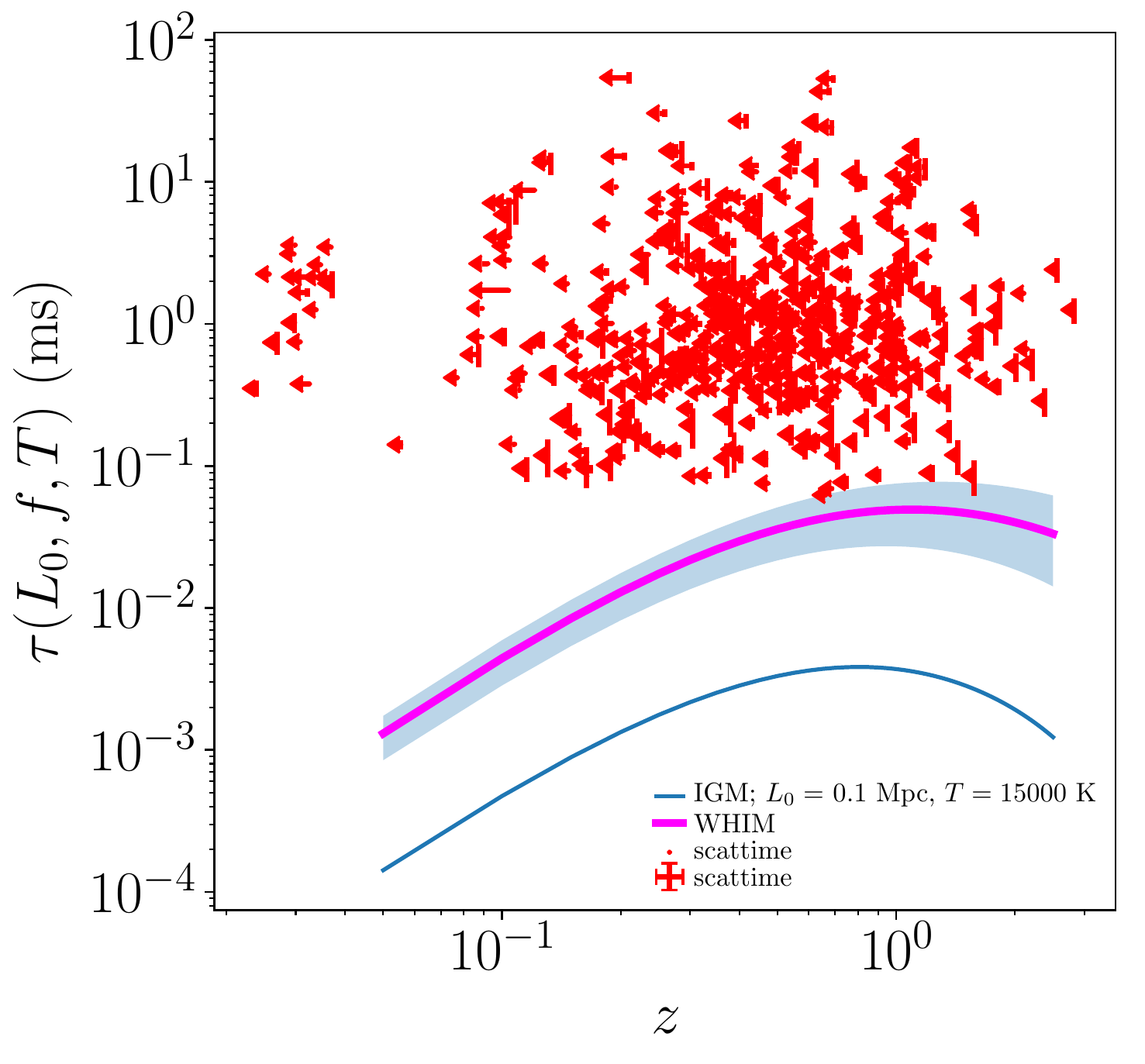}
    \caption{Intrinsic scattering timescales (denoted by ``scattime") as a function of inferred FRB redshift for the sources in the CHIME FRB catalog.  Overplotted are the IGM (blue thin line) and WHIM (thick magenta line) predictions, with the parameters as assumed in the main text.  We consider the scatter of the data points to result from contributions in the FRB sources and host galaxies. }
    \label{fig:scenarios}
\end{figure}

We use the First CHIME catalog of Fast Radio Bursts which comprises 536 transients, providing their Dispersion Measures (DMs), derived redshifts \citep{macquart2020}\footnote{The Transient Name Server (TNS, https://www.wis-tns.org/) is used to obtain the derived redshifts for the objects, which models the Galactic DM according to the NE2001 \citep{cordes2002} model. The redshifts $z$  here should be interpreted as the maximum redshifts $z_{\rm max}$ of the FRBs, keeping in mind the unknown host \citep{james2022} and Galactic halo contributions \citep{dolag2015, cook2023}.} and scattering timescales \citep{chime2021, chawla2022}. The distribution of intrinsic scattering timescales [related to the observed ones by $\tau = \tau_{\rm obs}/(1+z)$]  at the frequency of 600 MHz is plotted as the red points in Fig. \ref{fig:scenarios}.  The distribution has a fairly level lower envelope, which represents the minimum contribution of the IGM to the scattering timescale. The scattering can be modelled following  \citet{macquart2013} which connects the intrinsic timescale to that  associated with the turbulent material, by: 
\begin{eqnarray}
\tau &=& 
4.1 \times 10^{-5} \, (1+z_L)^{-1} \left( \frac{\lambda_0}{1\,{\rm m}} \right)^4 \left( \frac{D_{\rm eff}}{1\,{\rm Gpc} }\right) \nonumber \\
&& \left( \frac{{\rm SM}_{\rm eff}}{10^{12}\,{\rm m}^{-17/3} } \right) \left( \frac{l_0}{1\,{\rm AU}} \right)^{1/3} {\rm s} \nonumber \\
\label{tauEmpirical}
\end{eqnarray} 
in which 
$\lambda_0$ is the observed wavelength, $z_L$ is the assumed redshift of the turbulent material, and the ratio of  angular diameter distances is given by $D_{\rm eff} = D_L D_{LS}/D_S$. {The $D_L, D_{S}$ and $D_{LS}$ denote the angular diameter distances to $z_L$, to the source at $z$, and that between the source and turbulent material respectively.} We adopt the relation $z_L = \xi z$ with the fiducial value $\xi = 0.5$  throughout the analysis. The scattering measure is denoted by SM$_{\rm eff}$ and given by integrating the contribution of the IGM between the source and observer:
\begin{equation}
{\rm SM}_{\rm eff} = \int \frac{C_N^2 (l)}{(1+z')^2} dl = \int_0^{z} \frac{C_N^2(z') d_H(z')}{(1+z')^3} dz',
\label{scatmeas}
\end{equation}
 where $dl = d_H/(1+z) \  dz$ is the path length defined in terms of the Hubble distance $d_H = c/H(z)$,  {and $H(z) = H_0 (\Omega_m (1+z)^3 + \Omega_{\Lambda})^{1/2}$ is the Hubble parameter at redshift $z$.  We adopt a flat $\Lambda$CDM cosmology with the Hubble constant $H_0 = 71$ km/s/Mpc, and $\Omega_m = 0.3$ and $\Omega_{\Lambda}= 0.7$ being the ratios of the matter and dark energy densities to the critical density of the Universe,  respectively. } In the above equation,
 \begin{eqnarray}
C_N^2 (z) &=& \frac{\beta-3}{2 (2 \pi)^{4-\beta}} L_0^{3-\beta} f^2 \langle n_e(z) \rangle^2  \nonumber \\
&=& 9.42 \times 10^{-14}\, (1+z)^6\, f^2 \, \left(\frac{\Omega_b}{0.04} \right) \nonumber \\
 &\times &  \left( \frac{L_0}{1\,{\rm pc}} \right)^{-2/3} \,{\rm m}^{-20/3}.
  \label{CN2diffuse}
\end{eqnarray}
is amplitude of the turbulence per unit path length. The $C_N^2$ is expressed in terms of the quantity
$f^2 = (\langle n_e^2 \rangle - \langle n_e\rangle ^2)/\langle n_e \rangle^2 $, with the mean electron density set  equal to the  baryonic density of the Universe \citep{macquart2013}:
\begin{eqnarray}
\langle n_e \rangle = n_{e, 0} (1+z)^3 &=& \frac{\Omega_b \rho_{\rm crit}}{m_H} (1 - Y_{\rm He})(1 + 2 f_{\rm He}) (1+z)^3 \nonumber \\
&=&  2.26 \times 10^{-7} {\rm cm}^{-3} (1+z)^3
\end{eqnarray}
as is true for an almost completely ionized IGM.  {In the above,  $\rho_{\rm crit}$ is the critical matter density of the Universe,  $Y_{\rm He} = 0.24$ is the helium fraction and $f_{\rm He} = n_{\rm He}/n_H = 0.08$ \citep[e.g.,][]{munoz2018, macquart2020}.}
The clumping factor of the IGM is conventionally defined as  $C \equiv  f^2  + 1  \equiv \langle n_e^2\rangle/\langle n_e\rangle^2$.
The turbulence is assumed to follow a Kolmogorov spectrum {with $\beta$ being the power law index; $\beta = 11/3$} \citep{armstrong1995}. The outer scale $L_0$ is set to 0.1 Mpc in line with observations and theoretical estimates \citep{kunz2022}. 
We consider two approaches to constrain the lower scale  $l_0$: (i) as the Coulomb mean-free path in an ionized medium, and (ii) as the viscous scale of the intergalactic magnetic field. In the former case, the magnetic field is assumed non-existent, and $l_0$  is equal to the Coulomb mean-free path, expressed in terms of the plasma parameter $\Lambda$ given by:
\begin{equation}
\Lambda = \frac{4 \pi n_e \lambda_D^3}{3}
\end{equation}
where the Debye length is defined as (in cgs units):
\begin{eqnarray}
\lambda_D &=& \left(\frac{k_B T}{4 \pi n_e e^2}\right)^{1/2} \nonumber \\
&=& 2.2 \times 10^6 \  {\rm cm} \ \left(\frac{n_e}{10^{-7}  \ {\rm cm}^{-3}}\right)^{-1/2} \left(\frac{T}{10^4 \ {\rm K}}\right)^{1/2}
\end{eqnarray}
as a function of the ambient temperature $T$.
The mean free path is expressible in terms of $\Lambda$ and the Coulomb logarithm, $\ln \Lambda$ by:
\begin{equation}
 \lambda_{\rm mfp} = \frac{\Lambda}{\ln \Lambda} \lambda_{D}
\end{equation}

{{
\begin{table*}
\begin{center}
    \begin{tabular}{| c | c | c | c | c}
    \hline
 $z$ & $l_0$ (kpc)  & $B$ (nG) & $B_{\rm min} - B_{\rm max}$ (nG)  \\ \hline
1.05 & 0.07 & 11.8 & 5.8 - 26.9  \\
 1.3 & 0.08 &  14.3 & 7.0 - 32.3  \\
 1.55 & 0.09 & 17.9 & 8.6 - 40.1  \\
 1.8  & 0.12 & 22.8 & 10.9 - 50.8  \\
 2.05 &0.19 & 29.4 & 14.0 - 65.1  \\
\hline
    \end{tabular}
\end{center}
\caption{{ Upper limits on the intergalactic magnetic field assuming a photoionized IGM, at various redshifts (first column) and corresponding viscous scales $l_0$ (second column).  The third column denotes the derived field values assuming the outer scale $L_0 = 0.1$ Mpc. The range of possible  field values obtained on varying $L_0$ in an order of magnitude around 0.1 Mpc is indicated in the fourth column.}}
 \label{table:magfield}
\end{table*}
}}

\begin{figure}
    \centering
    \includegraphics[width = \columnwidth]{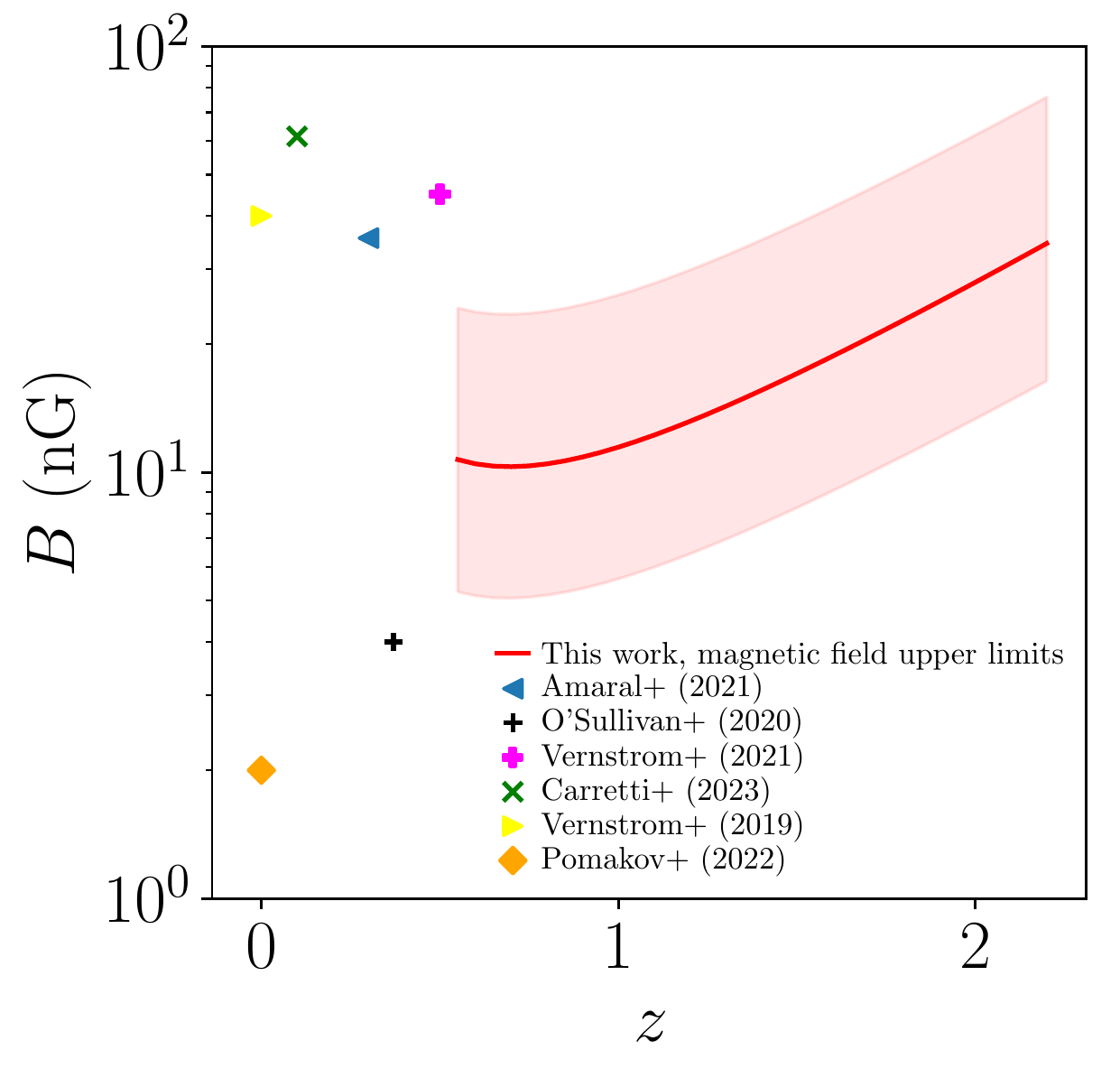} \includegraphics[width = \columnwidth]{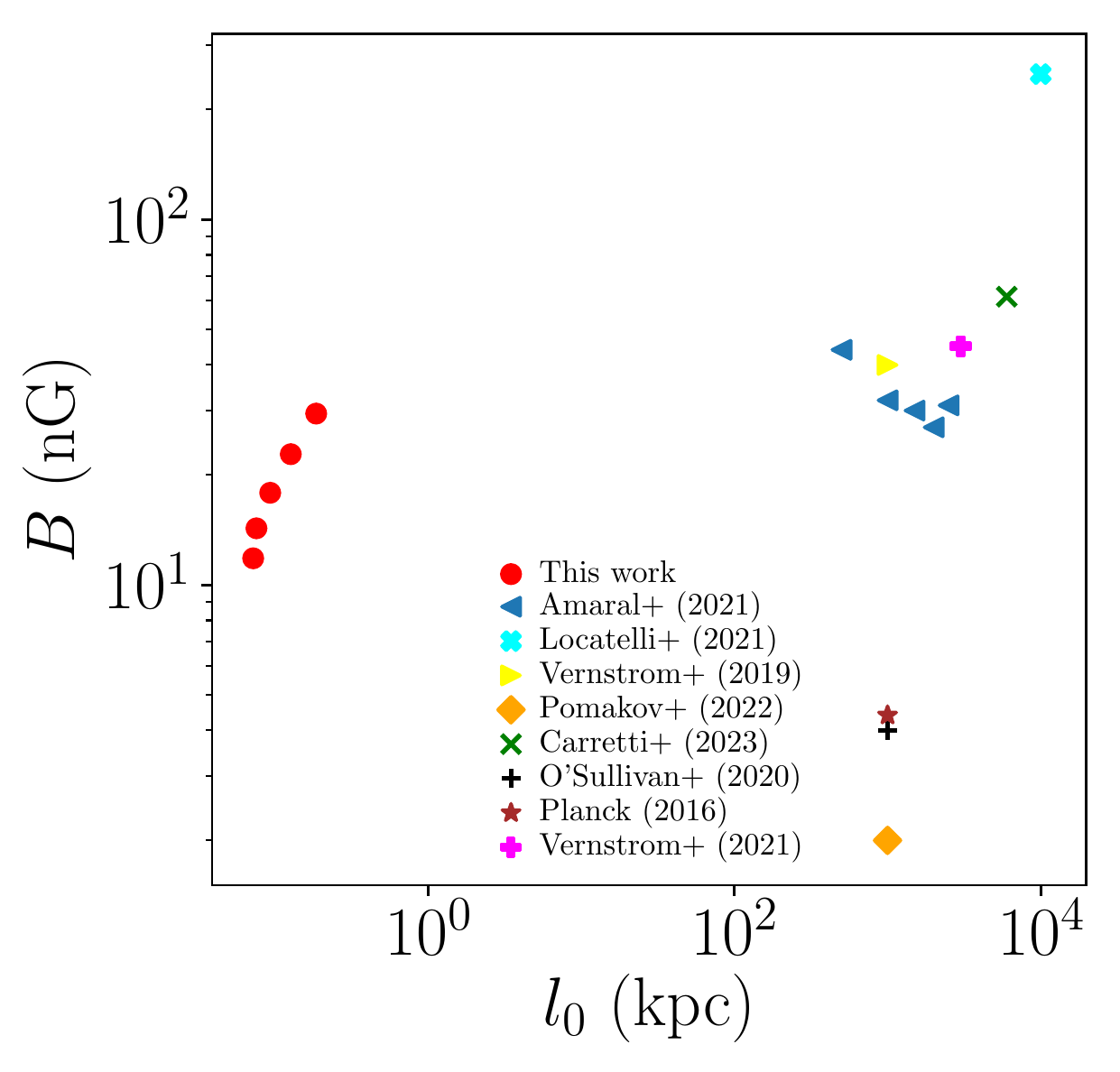}
    \caption{Upper limits on the magnetic field over $0 < z < 2$ and at the scales given in Table \ref{table:magfield}, as a function of (a) redshift and (b)  scale. Also plotted are existing upper limits from the literature over this redshift range \citep{vernstrom2019, vernstrom2021, pomakov2022, carretti2023, planck2016, locatelli2021, osullivan2020, amaral2021}. The shaded red area on the top panel indicates the range of $B$ allowed by plausible variations of the outer scale and velocity parameters, as described in the main text.}
    \label{fig:magfieldlimits}
\end{figure}
For the cases under consideration,  $\ln \Lambda \sim 30$. 

To estimate the smearing contribution from a nonmagnetized,  photoionized IGM between $z \sim 0-2$, { we  infer the expected clumping factor at each of the redshifts by the equation:
\begin{equation}
\Gamma_{\rm HI} \langle n_{\rm HI} \rangle = C_{\rm IGM} \langle n_e \rangle  \langle n_p \rangle \alpha_{\rm rec}
\end{equation}
}in which the photoionization rate is taken as $\Gamma_{\rm HI} = 10^{-12} s^{-1}$ from recent constraints \citep{mitra2018}, and we use the Case B recombination coefficient, $\alpha_{\rm rec} = 2.6 \times 10^{-13} \left(T_{\rm IGM}/10000 \ {\rm K} \right)^{-0.8} {\rm cm}^{3} {\rm s}^{-1}$ \citep{osterbrock1989} as is appropriate when considering average absorption in the IGM. We assume an average temperature $T_{\rm IGM} = 1.5 \times 10^4 $ K \citep{upton2016} at mean density, characteristic of the bulk of the IGM.
Assuming almost complete ionization (an excellent approximation in this regime) we can rewrite the above equation as:
\begin{equation}
\Gamma_{\rm HI} \langle f_{\rm HI} \rangle = C_{\rm IGM} \langle n_e \rangle  \alpha_{\rm rec}
\end{equation}
where $ f_{\rm HI} \equiv n_{\rm HI}/n_b = 10^{-5} h$ is the neutral fraction constrained by observations of the Lyman-$\alpha$ forest \citep{bi1993}, with $n_b \approx \langle n_e \rangle$ being the baryon number density.
Using $\langle n_e \rangle = 2.26 \times 10^{-7} {\rm cm}^{-3} (1+z)^3$, we find $C_{\rm IGM} \approx 167/(1+z)^3$ at $z \lesssim 3$
which, when used along with $T_{\rm IGM}$ in Eqs. (\ref{tauEmpirical}) and (\ref{CN2diffuse}),  leads to the blue line in Fig. \ref{fig:scenarios}. It can be seen that the predicted scattering timescales are well below the observations at all redshifts. We can thus conclude that a nonmagnetized, photoionized IGM is not sufficient to explain the observed smearing.

We now consider the contribution of a Warm-Hot Intergalactic Medium (WHIM) to the observed scattering. Between redshifts 0 and $\sim 0.6$, the WHIM occupies about 4 to 11 percent of the IGM by volume \citep[e.g.,][]{martizzi2019,danforth2016}. The clumping factor of the WHIM phase is about $C_{\rm WHIM} = 400$ \citep[e.g.,][]{dave2001}, with a characteristic temperature of $T_{\rm WHIM} = 5 \times 10^6$ K \citep{singari2020}. In the presence of the WHIM component, the effective temperature and clumping factor of the plasma become $\langle T \rangle = T_{\rm IGM} f_{\rm IGM} + T_{\rm WHIM} f_{\rm WHIM}; \langle C \rangle = C_{\rm IGM} f_{\rm IGM} + C_{\rm WHIM} f_{\rm WHIM}$ where $f_{\rm WHIM} = 0.04 - 0.11$ (all averages being over volume). Using these values in Eqs. (\ref{tauEmpirical}) and (\ref{CN2diffuse}) leads to the magenta line (bracketed by the shaded region covering the range of $f_{\rm WHIM}$) in Fig. \ref{fig:scenarios}. We find that while the WHIM can potentially provide the contribution to the smearing for the largest redshifts under consideration, $z \sim 1$, it cannot do so at lower redshifts, where a different source of turbulence may be warranted. 

\begin{figure}
    \centering
    \includegraphics[width = \columnwidth]{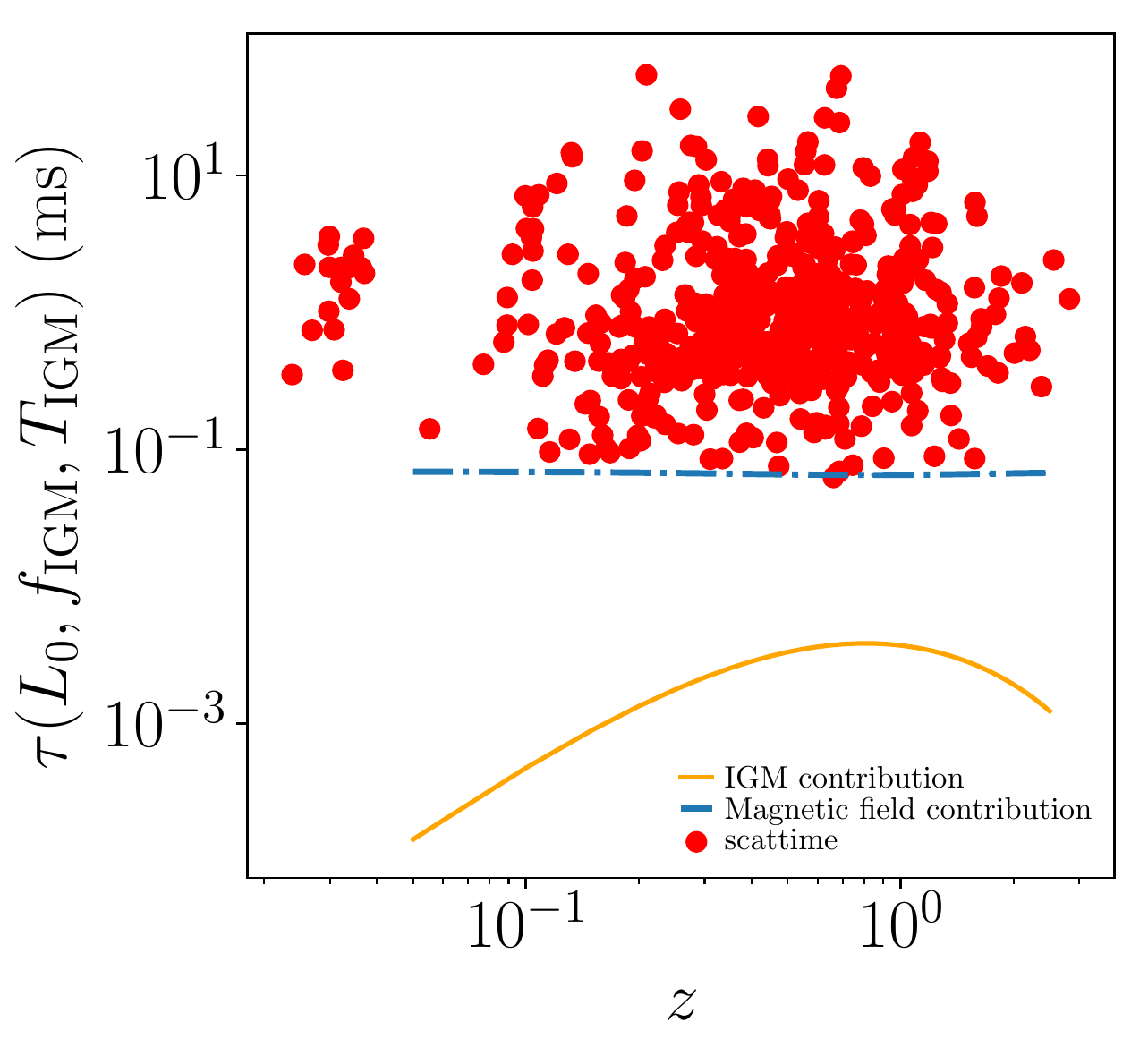}
    \caption{{{Inferred scattering timescale as a function of redshift  for a photoionized IGM including the inferred magnetic field $B$ in Table \ref{table:magfield}, shown by the blue dot-dashed line. This is compared to the CHIME observations (red data points).  The smearing timescale from a photoionized IGM alone (also plotted in Fig. \ref{fig:scenarios}) is shown as the orange solid line.}}}
    \label{fig:bestfits}
\end{figure}

We invoke a nonzero magnetic field in the IGM to account for the remainder of the turbulence.  { In this scenario, the  IGM is characterised by micro-instabilities resulting from growing Larmor-scale fluctuations that scatter ions, leading to an ``effective" viscosity with a  scale set by the Alf\'ven scale, also known as the effective viscous scale. This scale is the smallest energy scale at which eddies reside and is responsible for the turbulent energy cascade. Hence, it acts as the effective lower scale $l_0$ of turbulence in the presence of the  magnetic field $B$ \citep[e.g.,][]{kunz2022}}:
\begin{eqnarray}
\lambda_{\rm visc, B} &=& 4 \times 10^{-5} \  L_0 \left(\frac{U}{200  \ {\rm km/s}}\right)^{-3} \left(\frac{B}{1 \ {\rm n G}}\right)^{3} \nonumber \\
& \times & \left(\frac{n_e}{10^{-7}  \ {\rm cm}^{-3}}\right)^{-3/2}
\label{magvisc}
\end{eqnarray}
In the above,  $U$ is the associated velocity of the outer scale of turbulence, assumed to be 200 km/s (the typical circular velocity of virialized objects). 
{ We bin the $z \sim 0.75 - 2$ redshift range into equispaced redshift bins, and solve \eq{magvisc} for the strength of the field that accounts for the remainder of the smearing,  assuming the IGM contribution to be given by the solid line in Fig. \ref{fig:scenarios}.  We find that  values  of $10-30$ nG over $z \sim 1 - 2$ are needed to saturate the lower envelope of the scattering distribution,  as tabulated in the third column of Table \ref{table:magfield}.
The relevant inner scale is  found to be of the order of $0.07-0.20$ kpc over this redshift range, as shown in the second column of the Table. The constraints on the field strength are plotted in Fig. \ref{fig:magfieldlimits} along with other upper limits in the literature, both as a function of redshift, and  of scale.   The inferred $\tau$ values for the derived field strengths $B$ are plotted as a function of redshift in Fig. \ref{fig:bestfits} along with the data points.

If $L_0$ is allowed to vary in an order of magnitude around its fiducial value \citep{macquart2013, xu2020}, the constraints on $l_0$ go up to a few tens of kpc (since Eqs. \ref{scatmeas},\ref{CN2diffuse}  imply that for the same $\tau$, $l_0 \propto L_0^2$) while the magnetic field follows $B \propto L_0^{1/3}$. This range, indicated by the shaded  error band in the upper panel of Fig. \ref{fig:magfieldlimits} { and tabulated in the fourth column of Table \ref{table:magfield}}, is also representative of the allowed range of $B$ for a factor of $\sim 2$ variation in the assumed $U$ (since $B \propto U$ for fixed $l_0$ and $L_0$, \eq{magvisc}). 
}

\section{Summary}
We have used the measured distribution of scattering timescales from the CHIME catalog of 536 Fast Radio Bursts to place upper limits on the strength of the magnetic field on sub-kpc scales in the IGM. In so doing, we have introduced a new method to constrain IGM magnetic fields on much smaller scales than probed by existing techniques. A nonmagnetized, photoionized IGM is found insufficient to explain the observed smearing at all redshifts, with a WHIM component needed to marginally satisfy the constraints at $z \sim 1$. We find that magnetic fields of up to $\sim 10-30$ nG on scales of 0.07-0.20 kpc are allowed to account for the intergalactic component of the observed temporal smearing at $z \sim 1-2$. 
The constraints we obtain are close to those expected in models \citep{akahori2010, akahori2016, hackstein2020} requiring  larger simulated populations of FRBs. However, as seen in Fig. \ref{fig:magfieldlimits}, the scales probed are at least two orders of magnitude smaller, and the redshift ranges are higher, going up to $z \sim 2$. Our upper limits are also likely to be conservative, since a nonzero contribution to the turbulence is usually expected from the host galaxy, even for the sources at the lower 
envelope of the observed $\tau$ distribution. {\footnote{ The temporal resolution of the CHIME observations is not expected to affect the distribution of scattering timescales and derived results \citep[e.g.,][]{zhu2018}.  Some of the observed $\tau$ measurements are themselves upper limits, hence the actual inferred field is likely to be much lower than the values quoted here.}}

{ While we use a scattering screen positioned halfway between the Earth and the source in this paper,  it is worth exploring the effects of modifying this assumption  to account for nearer screens.  In the extreme case when the scattering screen is assumed to be located at the Laniakea supercluster ($z_L \sim 0.1$, $D_{\rm eff} \sim 0.33$ Gpc) for all the sources,  it results in a factor of $\sim 1.1 - 2$ higher $\tau$ compared to the blue and magenta curves in Fig. 1.  
This is not expected to have a significant effect on the derived magnetic field values at $z \gtrsim 1$, and does not affect their associated viscous scales, as well as the conclusion that the inferred values are upper limits.}

A potential independent constraint on these values may come from measurements of the RM and its scatter, $\sigma_{\rm RM}$ from the IGM alone. The present analysis predicts an RM contribution of $\sim$ 10 rad/m$^2$, and a scatter of $\sigma_{\rm RM} \sim 0.014 $ rad/m$^2$ from the magnetized IGM, consistently with expectations from simulations \citep[e.g.,][]{akahori2016}. These are about an order of magnitude lower than current observational constraints on RM and $\sigma_{\rm RM}$ \citep{feng2022,osullivan2019}, and thus likely to be subdominant to the host contribution, but may be observable with larger sample surveys. Future data from the Square Kilometre Array \citep[SKA;][]{gaensler2009} and its precursors can be used to further improve the magnetic field constraints on some of these smallest scales accessible   in the IGM.

\section*{Acknowledgements}  
 We thank Bryan Gaensler and Ruth Durrer for helpful comments on the manuscript, and the referee for a useful report that improved the presentation of the paper.
 HP's research is supported by the Swiss National Science Foundation via Ambizione Grant PZ00P2\_179934. The work of AL is supported in part by the Black Hole Initiative, which is funded by grants from the John Templeton Foundation and the Gordon and Betty Moore Foundation.

\def\aj{AJ}                   
\def\araa{ARA\&A}             
\def\apj{ApJ}                 
\def\apjl{ApJ}                
\def\apjs{ApJS}               
\def\ao{Appl.Optics}          
\def\apss{Ap\&SS}             
\def\aap{A\&A}                
\def\aapr{A\&A~Rev.}          
\def\aaps{A\&AS}              
\def\azh{AZh}                 
\def\baas{BAAS}
\def\jcap{JCAP}
\def\jrasc{JRASC}             
\def\memras{MmRAS}
\def\na{New Astronomy}
\def\nat{Nature}
\def\mnras{MNRAS}             
\def\pra{Phys.Rev.A}          
\def\prb{Phys.Rev.B}          
\def\prc{Phys.Rev.C}          
\def\prd{Phys.Rev.D}          
\def\prl{Phys.Rev.Lett}       
\def\pasp{PASP}               
\def\pasj{PASJ}
\def\physrep{Phys. Repts.}
\def\qjras{QJRAS}             
\def\skytel{S\&T}             
\def\solphys{Solar~Phys.}     
\def\sovast{Soviet~Ast.}      
\def\ssr{Space~Sci.Rev.}      
\def\zap{ZAp}                 
\let\astap=\aap
\let\apjlett=\apjl
\let\apjsupp=\apjs

\small{
\bibliographystyle{plainnat}
\bibliography{mybib}

\begin{thebibliography}{}
\expandafter\ifx\csname natexlab\endcsname\relax\def\natexlab#1{#1}\fi
\providecommand{\url}[1]{\href{#1}{#1}}
\providecommand{\dodoi}[1]{doi:~\href{http://doi.org/#1}{\nolinkurl{#1}}}
\providecommand{\doeprint}[1]{\href{http://ascl.net/#1}{\nolinkurl{http://ascl.net/#1}}}
\providecommand{\doarXiv}[1]{\href{https://arxiv.org/abs/#1}{\nolinkurl{https://arxiv.org/abs/#1}}}

\bibitem[{{Akahori} \& {Ryu}(2010)}]{akahori2010}
{Akahori}, T., \& {Ryu}, D. 2010, \apj, 723, 476,
  \dodoi{10.1088/0004-637X/723/1/476}

\bibitem[{{Akahori} {et~al.}(2016){Akahori}, {Ryu}, \&
  {Gaensler}}]{akahori2016}
{Akahori}, T., {Ryu}, D., \& {Gaensler}, B.~M. 2016, \apj, 824, 105,
  \dodoi{10.3847/0004-637X/824/2/105}

\bibitem[{{Amaral} {et~al.}(2021){Amaral}, {Vernstrom}, \&
  {Gaensler}}]{amaral2021}
{Amaral}, A.~D., {Vernstrom}, T., \& {Gaensler}, B.~M. 2021, \mnras, 503, 2913,
  \dodoi{10.1093/mnras/stab564}

\bibitem[{{Armstrong} {et~al.}(1995){Armstrong}, {Rickett}, \&
  {Spangler}}]{armstrong1995}
{Armstrong}, J.~W., {Rickett}, B.~J., \& {Spangler}, S.~R. 1995, \apj, 443,
  209, \dodoi{10.1086/175515}

\bibitem[{{Beck} {et~al.}(1996){Beck}, {Brandenburg}, {Moss}, {Shukurov}, \&
  {Sokoloff}}]{beck1996}
{Beck}, R., {Brandenburg}, A., {Moss}, D., {Shukurov}, A., \& {Sokoloff}, D.
  1996, \araa, 34, 155, \dodoi{10.1146/annurev.astro.34.1.155}

\bibitem[{{Bernet} {et~al.}(2008){Bernet}, {Miniati}, {Lilly}, {Kronberg}, \&
  {Dessauges-Zavadsky}}]{bernet2008}
{Bernet}, M.~L., {Miniati}, F., {Lilly}, S.~J., {Kronberg}, P.~P., \&
  {Dessauges-Zavadsky}, M. 2008, \nat, 454, 302, \dodoi{10.1038/nature07105}

\bibitem[{{Bi}(1993)}]{bi1993}
{Bi}, H. 1993, \apj, 405, 479, \dodoi{10.1086/172380}

\bibitem[{{Br{\"u}ggen} {et~al.}(2005){Br{\"u}ggen}, {Ruszkowski},
  {Simionescu}, {Hoeft}, \& {Dalla Vecchia}}]{bruggen2005}
{Br{\"u}ggen}, M., {Ruszkowski}, M., {Simionescu}, A., {Hoeft}, M., \& {Dalla
  Vecchia}, C. 2005, \apjl, 631, L21, \dodoi{10.1086/497004}

\bibitem[{{Carretti} {et~al.}(2023){Carretti}, {O'Sullivan}, {Vacca}, {Vazza},
  {Gheller}, {Vernstrom}, \& {Bonafede}}]{carretti2023}
{Carretti}, E., {O'Sullivan}, S.~P., {Vacca}, V., {et~al.} 2023, \mnras, 518,
  2273, \dodoi{10.1093/mnras/stac2966}

\bibitem[{{Cen} \& {Ostriker}(2006)}]{cen2006}
{Cen}, R., \& {Ostriker}, J.~P. 2006, \apj, 650, 560, \dodoi{10.1086/506505}

\bibitem[{{Chawla} {et~al.}(2022){Chawla}, {Kaspi}, {Ransom}, {Bhardwaj},
  {Boyle}, {Breitman}, {Cassanelli}, {Cubranic}, {Dong}, {Fonseca}, {Gaensler},
  {Giri}, {Josephy}, {Kaczmarek}, {Leung}, {Masui}, {Mena-Parra}, {Merryfield},
  {Michilli}, {M{\"u}nchmeyer}, {Ng}, {Patel}, {Pearlman}, {Petroff},
  {Pleunis}, {Rahman}, {Sanghavi}, {Shin}, {Smith}, {Stairs}, \&
  {Tendulkar}}]{chawla2022}
{Chawla}, P., {Kaspi}, V.~M., {Ransom}, S.~M., {et~al.} 2022, \apj, 927, 35,
  \dodoi{10.3847/1538-4357/ac49e1}

\bibitem[{{Cheng} {et~al.}(1996){Cheng}, {Olinto}, {Schramm}, \&
  {Truran}}]{cheng1996}
{Cheng}, B., {Olinto}, A.~V., {Schramm}, D.~N., \& {Truran}, J.~W. 1996, \prd,
  54, 4714, \dodoi{10.1103/PhysRevD.54.4714}

\bibitem[{{CHIME/FRB Collaboration} {et~al.}(2021){CHIME/FRB Collaboration},
  {Amiri}, {Andersen}, {Bandura}, {Berger}, {Bhardwaj}, {Boyce}, {Boyle},
  {Brar}, {Breitman}, {Cassanelli}, {Chawla}, {Chen}, {Cliche}, {Cook},
  {Cubranic}, {Curtin}, {Deng}, {Dobbs}, {Dong}, {Eadie}, {Fandino}, {Fonseca},
  {Gaensler}, {Giri}, {Good}, {Halpern}, {Hill}, {Hinshaw}, {Josephy},
  {Kaczmarek}, {Kader}, {Kania}, {Kaspi}, {Landecker}, {Lang}, {Leung}, {Li},
  {Lin}, {Masui}, {McKinven}, {Mena-Parra}, {Merryfield}, {Meyers}, {Michilli},
  {Milutinovic}, {Mirhosseini}, {M{\"u}nchmeyer}, {Naidu}, {Newburgh}, {Ng},
  {Patel}, {Pen}, {Petroff}, {Pinsonneault-Marotte}, {Pleunis},
  {Rafiei-Ravandi}, {Rahman}, {Ransom}, {Renard}, {Sanghavi}, {Scholz}, {Shaw},
  {Shin}, {Siegel}, {Sikora}, {Singh}, {Smith}, {Stairs}, {Tan}, {Tendulkar},
  {Vanderlinde}, {Wang}, {Wulf}, \& {Zwaniga}}]{chime2021}
{CHIME/FRB Collaboration}, {Amiri}, M., {Andersen}, B.~C., {et~al.} 2021,
  \apjs, 257, 59, \dodoi{10.3847/1538-4365/ac33ab}

\bibitem[{{Cook} {et~al.}(2023){Cook}, {Bhardwaj}, {Gaensler}, {Scholz},
  {Eadie}, {Hill}, {Kaspi}, {Masui}, {Curtin}, {Dong}, {Fonseca},
  {Herrera-Martin}, {Kaczmarek}, {Lanman}, {Lazda}, {Meyers}, {Michilli},
  {Pandhi}, {Pearlman}, {Pleunis}, {Ransom}, {Rahman}, {Sand}, {Shin}, {Smith},
  {Stairs}, \& {Stenning}}]{cook2023}
{Cook}, A.~M., {Bhardwaj}, M., {Gaensler}, B.~M., {et~al.} 2023, arXiv
  e-prints, arXiv:2301.03502.
\newblock \doarXiv{2301.03502}

\bibitem[{{Cordes} \& {Lazio}(2002)}]{cordes2002}
{Cordes}, J.~M., \& {Lazio}, T.~J.~W. 2002, arXiv e-prints, astro.
\newblock \doarXiv{astro-ph/0207156}

\bibitem[{{Danforth} {et~al.}(2016){Danforth}, {Keeney}, {Tilton}, {Shull},
  {Stocke}, {Stevans}, {Pieri}, {Savage}, {France}, {Syphers}, {Smith},
  {Green}, {Froning}, {Penton}, \& {Osterman}}]{danforth2016}
{Danforth}, C.~W., {Keeney}, B.~A., {Tilton}, E.~M., {et~al.} 2016, \apj, 817,
  111, \dodoi{10.3847/0004-637X/817/2/111}

\bibitem[{{Dav{\'e}} {et~al.}(2001){Dav{\'e}}, {Cen}, {Ostriker}, {Bryan},
  {Hernquist}, {Katz}, {Weinberg}, {Norman}, \& {O'Shea}}]{dave2001}
{Dav{\'e}}, R., {Cen}, R., {Ostriker}, J.~P., {et~al.} 2001, \apj, 552, 473,
  \dodoi{10.1086/320548}

\bibitem[{{Dolag} {et~al.}(1999){Dolag}, {Bartelmann}, \& {Lesch}}]{dolag1999}
{Dolag}, K., {Bartelmann}, M., \& {Lesch}, H. 1999, \aap, 348, 351.
\newblock \doarXiv{astro-ph/0202272}

\bibitem[{{Dolag} {et~al.}(2015){Dolag}, {Gaensler}, {Beck}, \&
  {Beck}}]{dolag2015}
{Dolag}, K., {Gaensler}, B.~M., {Beck}, A.~M., \& {Beck}, M.~C. 2015, \mnras,
  451, 4277, \dodoi{10.1093/mnras/stv1190}

\bibitem[{{Durrer} \& {Neronov}(2013)}]{durrer2013}
{Durrer}, R., \& {Neronov}, A. 2013, \aapr, 21, 62,
  \dodoi{10.1007/s00159-013-0062-7}

\bibitem[{{Feng} {et~al.}(2022){Feng}, {Li}, {Yang}, {Zhang}, {Zhu}, {Zhang},
  {Lu}, {Wang}, {Dai}, {Lynch}, {Yao}, {Jiang}, {Niu}, {Zhou}, {Xu}, {Miao},
  {Niu}, {Meng}, {Qian}, {Tsai}, {Wang}, {Xue}, {Yue}, {Yuan}, {Zhang}, \&
  {Zhang}}]{feng2022}
{Feng}, Y., {Li}, D., {Yang}, Y.-P., {et~al.} 2022, Science, 375, 1266,
  \dodoi{10.1126/science.abl7759}

\bibitem[{{Furlanetto} \& {Loeb}(2001)}]{furlanetto2001}
{Furlanetto}, S.~R., \& {Loeb}, A. 2001, \apj, 556, 619, \dodoi{10.1086/321630}

\bibitem[{{Gaensler}(2009)}]{gaensler2009}
{Gaensler}, B.~M. 2009, in Cosmic Magnetic Fields: From Planets, to Stars and
  Galaxies, ed. K.~G. {Strassmeier}, A.~G. {Kosovichev}, \& J.~E. {Beckman},
  Vol. 259, 645--652, \dodoi{10.1017/S1743921309031470}

\bibitem[{{Hackstein} {et~al.}(2020){Hackstein}, {Br{\"u}ggen}, {Vazza}, \&
  {Rodrigues}}]{hackstein2020}
{Hackstein}, S., {Br{\"u}ggen}, M., {Vazza}, F., \& {Rodrigues}, L.~F.~S. 2020,
  \mnras, 498, 4811, \dodoi{10.1093/mnras/staa2572}

\bibitem[{{James} {et~al.}(2022){James}, {Prochaska}, {Macquart},
  {North-Hickey}, {Bannister}, \& {Dunning}}]{james2022}
{James}, C.~W., {Prochaska}, J.~X., {Macquart}, J.~P., {et~al.} 2022, \mnras,
  509, 4775, \dodoi{10.1093/mnras/stab3051}

\bibitem[{Kahniashvili \& Ratra(2005)}]{Kahniashvili2005}
Kahniashvili, T., \& Ratra, B. 2005, Physical Review D, 71,
  \dodoi{10.1103/physrevd.71.103006}

\bibitem[{{Kawasaki} \& {Kusakabe}(2012)}]{kawasaki2012}
{Kawasaki}, M., \& {Kusakabe}, M. 2012, \prd, 86, 063003,
  \dodoi{10.1103/PhysRevD.86.063003}

\bibitem[{{Kunz} {et~al.}(2022){Kunz}, {Jones}, \& {Zhuravleva}}]{kunz2022}
{Kunz}, M.~W., {Jones}, T.~W., \& {Zhuravleva}, I. 2022, arXiv e-prints,
  arXiv:2205.02489.
\newblock \doarXiv{2205.02489}

\bibitem[{{Locatelli} {et~al.}(2021){Locatelli}, {Vazza}, {Bonafede}, {Banfi},
  {Bernardi}, {Gheller}, {Botteon}, \& {Shimwell}}]{locatelli2021}
{Locatelli}, N., {Vazza}, F., {Bonafede}, A., {et~al.} 2021, \aap, 652, A80,
  \dodoi{10.1051/0004-6361/202140526}

\bibitem[{{Lorimer} {et~al.}(2007){Lorimer}, {Bailes}, {McLaughlin},
  {Narkevic}, \& {Crawford}}]{lorimer2007}
{Lorimer}, D.~R., {Bailes}, M., {McLaughlin}, M.~A., {Narkevic}, D.~J., \&
  {Crawford}, F. 2007, Science, 318, 777, \dodoi{10.1126/science.1147532}

\bibitem[{{Macquart} \& {Koay}(2013)}]{macquart2013}
{Macquart}, J.-P., \& {Koay}, J.~Y. 2013, \apj, 776, 125,
  \dodoi{10.1088/0004-637X/776/2/125}

\bibitem[{{Macquart} {et~al.}(2020){Macquart}, {Prochaska}, {McQuinn},
  {Bannister}, {Bhandari}, {Day}, {Deller}, {Ekers}, {James}, {Marnoch},
  {Os{\l}owski}, {Phillips}, {Ryder}, {Scott}, {Shannon}, \&
  {Tejos}}]{macquart2020}
{Macquart}, J.~P., {Prochaska}, J.~X., {McQuinn}, M., {et~al.} 2020, \nat, 581,
  391, \dodoi{10.1038/s41586-020-2300-2}

\bibitem[{{Martizzi} {et~al.}(2019){Martizzi}, {Vogelsberger}, {Artale},
  {Haider}, {Torrey}, {Marinacci}, {Nelson}, {Pillepich}, {Weinberger},
  {Hernquist}, {Naiman}, \& {Springel}}]{martizzi2019}
{Martizzi}, D., {Vogelsberger}, M., {Artale}, M.~C., {et~al.} 2019, \mnras,
  486, 3766, \dodoi{10.1093/mnras/stz1106}

\bibitem[{{Mitra} {et~al.}(2018){Mitra}, {Choudhury}, \& {Ferrara}}]{mitra2018}
{Mitra}, S., {Choudhury}, T.~R., \& {Ferrara}, A. 2018, \mnras, 473, 1416,
  \dodoi{10.1093/mnras/stx2443}

\bibitem[{{Mu{\~n}oz} \& {Loeb}(2018)}]{munoz2018}
{Mu{\~n}oz}, J.~B., \& {Loeb}, A. 2018, \prd, 98, 103518,
  \dodoi{10.1103/PhysRevD.98.103518}

\bibitem[{{Osterbrock}(1989)}]{osterbrock1989}
{Osterbrock}, D.~E. 1989, {Astrophysics of gaseous nebulae and active galactic
  nuclei}

\bibitem[{{O'Sullivan} {et~al.}(2019){O'Sullivan}, {Machalski}, {Van Eck},
  {Heald}, {Br{\"u}ggen}, {Fynbo}, {Heintz}, {Lara-Lopez}, {Vacca},
  {Hardcastle}, {Shimwell}, {Tasse}, {Vazza}, {Andernach}, {Birkinshaw},
  {Haverkorn}, {Horellou}, {Williams}, {Harwood}, {Brunetti}, {Anderson},
  {Mao}, {Nikiel-Wroczy{\'n}ski}, {Takahashi}, {Carretti}, {Vernstrom}, {van
  Weeren}, {Orr{\'u}}, {Morabito}, \& {Callingham}}]{osullivan2019}
{O'Sullivan}, S.~P., {Machalski}, J., {Van Eck}, C.~L., {et~al.} 2019, \aap,
  622, A16, \dodoi{10.1051/0004-6361/201833832}

\bibitem[{{O'Sullivan} {et~al.}(2020){O'Sullivan}, {Br{\"u}ggen}, {Vazza},
  {Carretti}, {Locatelli}, {Stuardi}, {Vacca}, {Vernstrom}, {Heald},
  {Horellou}, {Shimwell}, {Hardcastle}, {Tasse}, \&
  {R{\"o}ttgering}}]{osullivan2020}
{O'Sullivan}, S.~P., {Br{\"u}ggen}, M., {Vazza}, F., {et~al.} 2020, \mnras,
  495, 2607, \dodoi{10.1093/mnras/staa1395}

\bibitem[{{Petroff} {et~al.}(2016){Petroff}, {Barr}, {Jameson}, {Keane},
  {Bailes}, {Kramer}, {Morello}, {Tabbara}, \& {van Straten}}]{petroff2016}
{Petroff}, E., {Barr}, E.~D., {Jameson}, A., {et~al.} 2016, \pasa, 33, e045,
  \dodoi{10.1017/pasa.2016.35}

\bibitem[{{Planck Collaboration} {et~al.}(2016){Planck Collaboration}, {Ade},
  {Aghanim}, {Arnaud}, {Arroja}, {Ashdown}, {Aumont}, {Baccigalupi},
  {Ballardini}, {Banday}, {Barreiro}, {Bartolo}, {Battaner}, {Benabed},
  {Beno{\^\i}t}, {Benoit-L{\'e}vy}, {Bernard}, {Bersanelli}, {Bielewicz},
  {Bock}, {Bonaldi}, {Bonavera}, {Bond}, {Borrill}, {Bouchet}, {Bucher},
  {Burigana}, {Butler}, {Calabrese}, {Cardoso}, {Catalano}, {Chamballu},
  {Chiang}, {Chluba}, {Christensen}, {Church}, {Clements}, {Colombi},
  {Colombo}, {Combet}, {Couchot}, {Coulais}, {Crill}, {Curto}, {Cuttaia},
  {Danese}, {Davies}, {Davis}, {de Bernardis}, {de Rosa}, {de Zotti},
  {Delabrouille}, {D{\'e}sert}, {Diego}, {Dolag}, {Dole}, {Donzelli},
  {Dor{\'e}}, {Douspis}, {Ducout}, {Dupac}, {Efstathiou}, {Elsner},
  {En{\ss}lin}, {Eriksen}, {Fergusson}, {Finelli}, {Florido}, {Forni},
  {Frailis}, {Fraisse}, {Franceschi}, {Frejsel}, {Galeotta}, {Galli}, {Ganga},
  {Giard}, {Giraud-H{\'e}raud}, {Gjerl{\o}w}, {Gonz{\'a}lez-Nuevo},
  {G{\'o}rski}, {Gratton}, {Gregorio}, {Gruppuso}, {Gudmundsson}, {Hansen},
  {Hanson}, {Harrison}, {Helou}, {Henrot-Versill{\'e}},
  {Hern{\'a}ndez-Monteagudo}, {Herranz}, {Hildebrandt}, {Hivon}, {Hobson},
  {Holmes}, {Hornstrup}, {Hovest}, {Huffenberger}, {Hurier}, {Jaffe}, {Jaffe},
  {Jones}, {Juvela}, {Keih{\"a}nen}, {Keskitalo}, {Kim}, {Kisner}, {Knoche},
  {Kunz}, {Kurki-Suonio}, {Lagache}, {L{\"a}hteenm{\"a}ki}, {Lamarre},
  {Lasenby}, {Lattanzi}, {Lawrence}, {Leahy}, {Leonardi}, {Lesgourgues},
  {Levrier}, {Liguori}, {Lilje}, {Linden-V{\o}rnle}, {L{\'o}pez-Caniego},
  {Lubin}, {Mac{\'\i}as-P{\'e}rez}, {Maggio}, {Maino}, {Mandolesi}, {Mangilli},
  {Maris}, {Martin}, {Mart{\'\i}nez-Gonz{\'a}lez}, {Masi}, {Matarrese},
  {McGehee}, {Meinhold}, {Melchiorri}, {Mendes}, {Mennella}, {Migliaccio},
  {Mitra}, {Miville-Desch{\^e}nes}, {Molinari}, {Moneti}, {Montier},
  {Morgante}, {Mortlock}, {Moss}, {Munshi}, {Murphy}, {Naselsky}, {Nati},
  {Natoli}, {Netterfield}, {N{\o}rgaard-Nielsen}, {Noviello}, {Novikov},
  {Novikov}, {Oppermann}, {Oxborrow}, {Paci}, {Pagano}, {Pajot}, {Paoletti},
  {Pasian}, {Patanchon}, {Perdereau}, {Perotto}, {Perrotta}, {Pettorino},
  {Piacentini}, {Piat}, {Pierpaoli}, {Pietrobon}, {Plaszczynski},
  {Pointecouteau}, {Polenta}, {Popa}, {Pratt}, {Pr{\'e}zeau}, {Prunet},
  {Puget}, {Rachen}, {Rebolo}, {Reinecke}, {Remazeilles}, {Renault}, {Renzi},
  {Ristorcelli}, {Rocha}, {Rosset}, {Rossetti}, {Roudier},
  {Rubi{\~n}o-Mart{\'\i}n}, {Ruiz-Granados}, {Rusholme}, {Sandri}, {Santos},
  {Savelainen}, {Savini}, {Scott}, {Seiffert}, {Shellard}, {Shiraishi},
  {Spencer}, {Stolyarov}, {Stompor}, {Sudiwala}, {Sunyaev}, {Sutton},
  {Suur-Uski}, {Sygnet}, {Tauber}, {Terenzi}, {Toffolatti}, {Tomasi},
  {Tristram}, {Tucci}, {Tuovinen}, {Umana}, {Valenziano}, {Valiviita}, {Van
  Tent}, {Vielva}, {Villa}, {Wade}, {Wandelt}, {Wehus}, {Yvon}, {Zacchei}, \&
  {Zonca}}]{planck2016}
{Planck Collaboration}, {Ade}, P.~A.~R., {Aghanim}, N., {et~al.} 2016, \aap,
  594, A19, \dodoi{10.1051/0004-6361/201525821}

\bibitem[{{Pomakov} {et~al.}(2022){Pomakov}, {O'Sullivan}, {Br{\"u}ggen},
  {Vazza}, {Carretti}, {Heald}, {Horellou}, {Shimwell}, {Shulevski}, \&
  {Vernstrom}}]{pomakov2022}
{Pomakov}, V.~P., {O'Sullivan}, S.~P., {Br{\"u}ggen}, M., {et~al.} 2022,
  \mnras, 515, 256, \dodoi{10.1093/mnras/stac1805}

\bibitem[{{Pshirkov} {et~al.}(2016){Pshirkov}, {Tinyakov}, \&
  {Urban}}]{pshirkov2016}
{Pshirkov}, M.~S., {Tinyakov}, P.~G., \& {Urban}, F.~R. 2016, \prl, 116,
  191302, \dodoi{10.1103/PhysRevLett.116.191302}

\bibitem[{Quashnock {et~al.}(1989)Quashnock, Loeb, \& Spergel}]{Quashnock1988}
Quashnock, J.~M., Loeb, A., \& Spergel, D.~N. 1989, Astrophys.J., 344, L49,
  \dodoi{10.1086/185528}

\bibitem[{{Ravi} {et~al.}(2016){Ravi}, {Shannon}, {Bailes}, {Bannister},
  {Bhandari}, {Bhat}, {Burke-Spolaor}, {Caleb}, {Flynn}, {Jameson}, {Johnston},
  {Keane}, {Kerr}, {Tiburzi}, {Tuntsov}, \& {Vedantham}}]{ravi2016}
{Ravi}, V., {Shannon}, R.~M., {Bailes}, M., {et~al.} 2016, Science, 354, 1249,
  \dodoi{10.1126/science.aaf6807}

\bibitem[{{Ryu} {et~al.}(2008){Ryu}, {Kang}, {Cho}, \& {Das}}]{ryu2008}
{Ryu}, D., {Kang}, H., {Cho}, J., \& {Das}, S. 2008, Science, 320, 909,
  \dodoi{10.1126/science.1154923}

\bibitem[{{Singari} {et~al.}(2020){Singari}, {Ghosh}, \&
  {Khatri}}]{singari2020}
{Singari}, B., {Ghosh}, T., \& {Khatri}, R. 2020, \jcap, 2020, 028,
  \dodoi{10.1088/1475-7516/2020/08/028}

\bibitem[{{Subramanian}(2016)}]{subramanian2016}
{Subramanian}, K. 2016, Reports on Progress in Physics, 79, 076901,
  \dodoi{10.1088/0034-4885/79/7/076901}

\bibitem[{{Thornton} {et~al.}(2013){Thornton}, {Stappers}, {Bailes},
  {Barsdell}, {Bates}, {Bhat}, {Burgay}, {Burke-Spolaor}, {Champion}, {Coster},
  {D'Amico}, {Jameson}, {Johnston}, {Keith}, {Kramer}, {Levin}, {Milia}, {Ng},
  {Possenti}, \& {van Straten}}]{thornton2013}
{Thornton}, D., {Stappers}, B., {Bailes}, M., {et~al.} 2013, Science, 341, 53,
  \dodoi{10.1126/science.1236789}

\bibitem[{{Upton Sanderbeck} {et~al.}(2016){Upton Sanderbeck}, {D'Aloisio}, \&
  {McQuinn}}]{upton2016}
{Upton Sanderbeck}, P.~R., {D'Aloisio}, A., \& {McQuinn}, M.~J. 2016, \mnras,
  460, 1885, \dodoi{10.1093/mnras/stw1117}

\bibitem[{{Vall{\'e}e}(2004)}]{vallee2004}
{Vall{\'e}e}, J.~P. 2004, \nar, 48, 763, \dodoi{10.1016/j.newar.2004.03.017}

\bibitem[{{Vazza} {et~al.}(2017){Vazza}, {Br{\"u}ggen}, {Gheller}, {Hackstein},
  {Wittor}, \& {Hinz}}]{vazza2017}
{Vazza}, F., {Br{\"u}ggen}, M., {Gheller}, C., {et~al.} 2017, Classical and
  Quantum Gravity, 34, 234001, \dodoi{10.1088/1361-6382/aa8e60}

\bibitem[{{Vazza} {et~al.}(2014){Vazza}, {Br{\"u}ggen}, {Gheller}, \&
  {Wang}}]{vazza2014}
{Vazza}, F., {Br{\"u}ggen}, M., {Gheller}, C., \& {Wang}, P. 2014, \mnras, 445,
  3706, \dodoi{10.1093/mnras/stu1896}

\bibitem[{{Vernstrom} {et~al.}(2017){Vernstrom}, {Gaensler}, {Brown}, {Lenc},
  \& {Norris}}]{vernstrom2017}
{Vernstrom}, T., {Gaensler}, B.~M., {Brown}, S., {Lenc}, E., \& {Norris}, R.~P.
  2017, \mnras, 467, 4914, \dodoi{10.1093/mnras/stx424}

\bibitem[{{Vernstrom} {et~al.}(2019){Vernstrom}, {Gaensler}, {Rudnick}, \&
  {Andernach}}]{vernstrom2019}
{Vernstrom}, T., {Gaensler}, B.~M., {Rudnick}, L., \& {Andernach}, H. 2019,
  \apj, 878, 92, \dodoi{10.3847/1538-4357/ab1f83}

\bibitem[{{Vernstrom} {et~al.}(2021){Vernstrom}, {Heald}, {Vazza}, {Galvin},
  {West}, {Locatelli}, {Fornengo}, \& {Pinetti}}]{vernstrom2021}
{Vernstrom}, T., {Heald}, G., {Vazza}, F., {et~al.} 2021, \mnras, 505, 4178,
  \dodoi{10.1093/mnras/stab1301}

\bibitem[{{Xu} \& {Zhang}(2020)}]{xu2020}
{Xu}, S., \& {Zhang}, B. 2020, \apjl, 898, L48,
  \dodoi{10.3847/2041-8213/aba760}

\bibitem[{{Zhu} {et~al.}(2018){Zhu}, {Feng}, \& {Zhang}}]{zhu2018}
{Zhu}, W., {Feng}, L.-L., \& {Zhang}, F. 2018, \apj, 865, 147,
  \dodoi{10.3847/1538-4357/aadbb0}

\end{thebibliography}
}

\end{document}